**Driven Translocation of a Polynucleotide Chain Through a Nanopore—A Continuous Time Monte-Carlo Study**

Pui-Man Lam[1,3], Fei Liu[2] and Zhong-can Ou-Yang[3]

[1]Physics Department, Southern University, Baton Rouge, Louisiana 70813
[2]Center for Advanced Study, Tsinghua University, Beijing 100084, China
[3]Institute of Theoretical Physics, The Chinese Academy of Sciences, P.O. Box 2735
Beijing 100080, China

Using continuous time Monte-Carlo method we simulated the translocation of a polynucleotide chain driven through a nanopore by an electric field. We have used two models of driven diffusion due to the electric field. The chain may have strong interaction with the pore, and depends on which end of the chain first enters the pore. Depending on this interaction, in both cases, the distribution of times for the chain to pass through the pore in our model is found to have three peaks, as observed in the experiment of Kasianowicz, Brandin, Branton and Deamer (KBBD).

## I. INTRODUCTION

Recently there has been a lot of interest in the problem of the translocation of biopolymers chains driven through a nanopore by an electric field [ 1-12 ]. Such pores are 1-2 nm in size and would allow single-stranded but not double stranded DNA to pass through. The process of translocation of biopolymers through pores in membranes is ubiquitous in cell biology since most cells must transport macromolecules across membranes to function. Transcribed mRNA molecules for example are transported out of the nucleus through a nuclear pore complex. Viral injection of DNA into a host cell is another example. It has also the potential to be used as single-molecule tool and may eventually lead to a single-molecule RNA and DNA sequencing technique. For instance, Gerland et al [3] investigated the theoretical possibility of utilizing polymer translocation to determine the full basepairing pattern of polynucleotides, including RNA pseudoknots. Besides nanopores in biomembranes, one has also studied polymer translocation through solid-state nanopores [11-14].

Kasianowicz et al. [1] (KBBD) showed that an electric field can be used to drive single-stranded polynucleotides (poly[U]) molecules through an ionic channel in a lipid bilayer membrane. The pore was 1.5 nm in diameter at its narrowest constriction, barely larger than the diameter of a single polynucleotide strand. Single stranded, homogeneous, polynucleotides (poly[U]), close to monodisperse of 210 monomers in length were introduced into one side of the membrane, called the cis side. After applying a transmembrane potential of between 110 and 140 mV KBBD monitored the transmembrane ionic current as a function of time. This ionic current was almost constant, except for certain periods on the order of hundreds of microseconds, in which the current decreased by more than 90%. These periods of very low current were the times when a polynucleotide was in the process of passing through the pore and therefore blocking the current. They can thus be interpreted as giving the measurements of the times required for individual polynucleotides to transverse the membrane under the influence of an electric field. When the number of observed blockades were plotted against the length or lifetime of the blockades, one could clearly see three distinct peaks. The first peak corresponding to the shortest lifetime was found to be independent of the polymer length or applied potential. They reasoned that this peak was caused by polymers that entered and retracted and thus did not completely cross the membrane. On the other hand the mean lifetime of the other two peaks were directly related to polymer length and inversely related to applied potential and were both thus caused by polymers actually passing through the pore. The charge on each nucleotide is just the electron charge e. Using 125 mV for the electric potential, this gives $eV \approx 5k_BT$, for the electrostatic energy gained by moving one nucleotide completely through the pore, where $k_B$ is the Boltzman constant and T is the absolute temperature. KBBD made the intriguing suggestion that there are two characteristic times associated with translocation because the polynucleotide can enter the pore in two distinct directions: One peak corresponds to polymers that enter the channel with their 3' end first, the other to polymers that enter with their 5' end first.

Lubensky and Nelson [2] studied theoretically the polymer translocation problem in the experiment of KBBD. The polymer being constrained to pass through a tiny nanopore makes it a one-dimensional problem. They studied the probability P(x,t) that a contour length x of the polymer's backbone has passed through the pore at time t. Assuming that the probability current j defined by $\partial P / \partial t + \partial j / \partial x = 0$, to be proportional to P and to $\partial P / \partial x$, i.e.

$$j(x,t) = vP(x,t) - D\frac{\partial P(x,t)}{\partial x} \quad , \qquad (1)$$

they obtained the familiar equation for diffusion with drift

$$\frac{\partial P}{\partial t} = D\frac{\partial^2 P}{\partial x^2} - v\frac{\partial P}{\partial x} \qquad (2)$$

where v and D are, respectively, and average drift velocity and an effective diffusion coefficient. The solution of (2), subject to the boundary conditions that P vanish at x=0 and x=L, where L is the length of the polymer, and the initial condition $P(x,t=0) = d(x - x_0)$, can be expressed as an infinite series in terms of the eigenfunctions of the linear differential operator on the right hand side of (2). The probability that the polymer will exit the channel at x=L at time t is given by $j(t) = j(L)$, which however, is a very slowly converging infinite series. Fortunately, by using the Poisson resummation formual, it can be converted into another infinite series that is so rapidly converging that it is sufficient to take only the lowest order term. At this point a pathology in the model appeared: the starting point $x_0$ in the polymer cannot be taken to be zero, which is the case of interest. In the limit $x_0 \to 0$, the probability that the polymer passes through the pore, given by $c(x_0) = \int_0^\infty j(t)dt$ vanishes i.e. c(0)=0. Meaningful result can only be obtained by normalizing $j(t)$ by the total probability of passage, i.e. by defining the first passage probability as $y(t) = \lim_{x_0 \to 0} j(t)/c(x_0)$. For given v and D, the probability $y(t)$ that the polynucleotide takes a time t to pass through the channel has only one peak. It is quite skewed and its mean and maximum are correspondingly well separated and is visibly different from a Gaussian with the same mean and variance. To explain the presence of two peaks in the data of KBBD, Lubensky and Nelson suggest that due to the strong interaction of the polymer with the pore, it is indeed possible that a polynucleotide passing through the pore with its 3’ end first can have an average velocity that is significantly different from one passing through with its 5’ end first. They proposed an interaction u(x) for the interaction of the polymer with the pore of the sawtooth form:

$$u(x) = \frac{u_0}{a}\frac{x}{b} \ , \quad x \le ab \qquad (3a)$$

$$u(x) = \frac{u_0}{1-a}\frac{b-x}{b} \ , \quad x > ab \qquad (3b)$$

where $u_0$ is a constant amplitude and $a$ is an asymmetry parameter with the symmetric case given by $a = 1/2$, whereas $a = 0$ or 1 correspond to maximal asymmetry. This potential is periodic with period b which is the distance between nucleotides. Lubensky and Nelson suggested that with this asymmetric interaction between the polymer and the pore, the effective mobility and diffusion constant of the polymer through the pore could take different values depending on whether the polymer entered the pore with the 3’ or the 5’ end first. They did not show however that this could indeed lead to two peaks in the distribution of passage times as in the experimental data. One may contemplate a numerical solution of the driven diffusion equation corresponding to (2), taking into account the potential (3). However, due to the aforementioned pathology of the model, at least in the special case $u_0 = 0$, resulting in the necessity of normalizing the passage probability by dividing with the total passage c(0), such a numerical procedure may be difficult to carry out. For this reason we have resort to a different procedure, the continuous time Monte-Carlo method to study the driven diffusion of a polymer through a nanopore taking into account the asymmetric interaction of the polymer with pore, in order to see if this indeed leads to the appearance of two peaks in the distribution of passage times. In section II we present the continuous time Monte-Carlo method, its application to the present problem and the results we obtained. Section III is the conclusion and discussion.

## II. CONTINUOUS TIME MONTE-CARLO METHOD

As a variant of the standard Monte Carlo method, the continuous time Monte Carlo [15, 16](CTMC) method is very efficient and fast because of the lack of waiting times due to rejection. In contrast to standard MC method, instead of the MC step used to approximate the real time, the “time” in Gillespie’s method could be the real physical time if the transition probabilities were calculated by first principles or empirically.

We first consider the case where there is no interaction between the polymer and the pore. The membrane with the pore separate the system into two parts, the cis side on the left where the polynucleotide is originally located, and the trans side on the right to which it will eventually translocate to. The membrane can be assumed to be perpendicular to the x-axis, with the pore at the position x=0. We assume that both ends of the polymer are right next to the pore on the cis side to start with and one end, the right end, is driven through the pore by an applied external electric field in the x direction, with the other end, the left end, staying always next to the pore until the whole polymer has passed through. This assumption simplifies the calculation but has no effect on the final result as we check afterwards using different contour lengths for the polymer. If the polymer consists of n nucleotides, its contour length will be nb, where b=0.56 nm is the length of a single nucleotide. Let x denote the projection of the right end of the polymer on the trans side on the x-axis. Then the end-to-end distance of the polymer is x since the left end is at the position x=0. In the freely-jointed-chain approximation, the free energy is given by

$$W(x,n) = W_0(x,n) + W_1(x,n) \tag{4}$$

with

$$W_0(x,n) = \frac{l k_B T}{b} x \tag{5}$$

$$W_1(x,n) = x f(x,n) - \int_0^{f(x,n)} x(f',n)\, df' \tag{6}$$

where the extension x is given by the Langevin function

$$x(f',n) = nb\left[\coth\left(\frac{f'P}{k_B T}\right) - \frac{k_B T}{f'P}\right] \tag{7}$$

with P=1.5 nm, the persistence length of the polymer.

We can now simulate diffusion using CTMC by changing randomly $x \to x \pm d$ , with $d = 0.1$ nm and calculating the transition rates [17, 18] from transition state theory

$$k_1 = \frac{1}{t_0} \exp\left( \frac{W(x) - W(x+d)}{k_B T} \right) \quad (8a)$$

$$k_2 = \frac{1}{t_0} \exp\left( \frac{W(x) - W(x-d)}{k_B T} \right) \quad (8b)$$

where $t_0^{-1}$ is an attempt frequency to be determined later. In continuous time Monte Carlo method, the acceptance of a chosen process is always set to one. In this way there is no rejection as in standard Monte Carlo method. However the choice of a given process is dictated by the rates. From $k_1, k_2$ we can define the probabilities

$$p_1 = \frac{k_1}{k_1 + k_2} \quad (9a)$$

$$p_2 = \frac{k_2}{k_1 + k_2} \quad (9b)$$

Then by generating two random numbers $g_1, g_2 \leq 1$, we can choose the new configuration j by the condition

$$g_1 \leq \sum_i^j p_i \quad (10)$$

The time is now incremented by the amount

$$\Delta t = -\frac{1}{k_1 + k_2} \log g_2 \quad (11)$$

Note that from (8), $p_1, p_2$ are independent of $t_0$, so that from (10), by choosing $t_0 = 1$, the time will then be in units of $t_0$. Since the transition rates and probabilities are clearly physically motivated, the calculated time should be the physical time.

We will first study the case with W=$W_0$, which is the case studied by Lubensky and Nelson, when there is no interaction between the polymer and the pore. The Langevin equation giving the time dependence of x can be obtained from (2)

$$\frac{dx}{dt} = v + h(t) \qquad (12)$$

where $h(t)$ is white noise with correlation $< h(t)h(t') >= 2Dd(t-t')$. The solution is

$$x = vt + \int_0^t h(t')dt'. \qquad (13)$$

The transit time $t_p$ corresponds to $x = L$, which gives

$$t_p = \frac{L}{v} - \frac{1}{v}\int_0^{t_p} h(t')dt' \qquad (14)$$

Since $h(t)$ is Gaussian random noise, this shows that $t_p$ is Gaussian distributed about the average value L/v. One can also calculate the average fluctuation of $t_p$ about its average value $(dt_p)^2 = \left\langle \left( t_p - \frac{L}{v} \right)^2 \right\rangle$, using the correlation of the random noise $h(t)$.

One easily find that $(dt_p)^2 = 2Dt_p$.

In Figure 1a we show our simulation result of the distribution of first passage times for different polymer lengths Nb, with N=50, 100 and 150, using $l = 5$ in $W_0$.

For each N, the first peak at the very left corresponds to cases when the polymer partially enter the trans side but then retracted into the cis side. This peak is independent of the polymer length L=Nb. The second peak at a larger lifetime is Gaussian in shape and corresponds to cases when the polymer actually transmitted through to the trans side. The lifetimes corresponding to this peak is proportional to the length of the polymer. The transit time in our simulation is obtained by monitoring the time when the length of polymer transmitted x is equal to L. Our result for the transit time distribution is in agreement with the result of the Langevin equation.

This is in agreement with our results presented in Fig.1a but in disagreement with those of Lubensky and Nelson obtained using a different definition of the first passage time distribution. In Fig. 1b we plot the average fluctuation $dt_p$ versus $t_p^{1/2}$. The result is a straight line, also confirming the result of the Langevin equation.

Next, we include also the free energy due to stretching of the polymer, i.e. $W = W_0 + W_1$. The results are presented in Fig. 2, again using $l = 5$ in $W_0$. The results are similar to those of Fig.1, except that the passage times are now larger due to the presence of the stretching term.

Now we study the case where there is an interaction between the polymer and the pore, i.e. we use $W = W_0 + W_1 + u(x)$, where u(x) is that given in (3). By varying the parameters $l$ in $W_0$, $u_0$ and $a$ in u(x), we readily obtain three distinct peaks in the transit time distribution. However, the distributions look quite different from the experimental data of KBBD. In Fig. 3 we show the distribution for the case $l = 5$ in $W_0$, $u_0 = 0.2k_BT$ and $a = 0.1$ in u(x), using N=210.

We have used a free energy $W_0 = -l\,k_B T x / b$ due to the electric field, which give rise to a constant force pulling on the polymer through the pore. Since this does not give good agreement with the experimental data of KBBD, we want to try another form of the free energy $W_0'(x) = -\frac{l\,k_B T}{b^2}\frac{x^2}{2}$. The charge on each nucleotide is the electronic charge e. If a length x'of the polymer has passed through the pore, the number of nucleotides having passed through is x'/b. In an external electric field E, the force pulling at the pore from the trans side will be eEx'/b. The work in pulling a length x through the pore is the integral $\int_0^x eE\frac{x'}{b}dx' = \frac{eEx^2}{2b} \equiv \frac{l' k_B T}{b^2}\frac{x^2}{2}$. The force due to the electric field on the polymer on the cis side will be counteracted and cancelled by the membrane. Such a model would not be unreasonable. Of course the average passage would now no longer be proportional to the length of the polymer.

We will first study the case with only the term due to the external electric field $W(x) = W_0^{'}(x)$. In that case the probability distribution P(x,t) satisfies a drift diffusion equation similar to (2):

$$\frac{\partial P}{\partial t} = D\frac{\partial^2 P}{\partial x^2} - \frac{\partial}{\partial x}\left(\frac{x}{t_f}P\right) \qquad (15)$$

where $t_f$ is a constant characteristic time due to the external field. The Langevin equation giving the time dependence of x is given by

$$\frac{dx}{dt} = \frac{x}{t_f} + h(t) \qquad (16)$$

where $h(t)$ is white noise with correlation $< h(t)h(t') >= 2Dd(t-t')$. The solution of this is

$$x = \exp(t/t_f)\int_0^t \exp(-t'/t_f)h(t')dt' \qquad (17)$$

The transit time $t_p$ corresponds to $x = L$, which gives

$$t_p = t_f \log L - t_f \log \int_0^{t_p} \exp(-t'/t_f)h(t')dt' \qquad (18)$$

Now due to the nonlinear logarithm dependence in the noise, the transit time is no longer Gaussian distributed and its average value cannot be easily evaluated. But the average transit time dependence on the polymer length can at most be logL. In fact due to the dependence of the upper limit of the integral on $t_p$ itself, the average transit time can actually saturate for large L and this is what we found in our simulation. Similarly the dependence of $t_p$ on the $t_f$ is linear for small L and this is what we find also in our simulation.

In Fig. 4 we show the distribution of transit times for $l^{'} = 3$, N=200. The main peak in the distribution looks indeed non-Gaussian, as predicted. We also find that the distribution is insensitive to N for N>20. So the average transit time actually saturates for N>20. In Fig. 5 we show the average transit time versus $1/l^{'}$ which is proportional to the characteristic time $t_f$ due to the external field. It shows indeed that the average transit time is proportional to $t_f$, as predicted.

We have now confirmed the agreement of our simulation result with the that of the predictions of the Langevin equation, in the case of the free energy $W(x) = W_0^{'}(x)$.

We can now proceed with the simulation for the case where we include also the stretching of the polymer in the free energy, i.e. $W(x) = W_0^{'}(x) + W_1(x)$. We find that in this case the transit time distribution is almost identical to the case with $W(x) = W_0^{'}(x)$. The effect of the polymer stretching has no effect in this case.

We then consider the case when there is an interaction between the polymer and the pore by simply adding the interaction potential u(x) to the free energy so that $W(x) = W_0^{'}(x) + W_1(x) + u(x)$. We first tried an interaction potential given in (3) as suggested by Lubensky and Nelson. However, in this case we were not able to obtain three visibly distinct peaks in the distribution of lifetimes, by adjusting the parameters $u_0, a$ and also $l^{'}$ in $W_0^{'}(x)$. Therefore we tried a different interaction potential of the following form: $u(x) = u_0 x / b$, when the polymer enters the pore with one end first and $u(x) = u_0'(b - x)/b$, when it enters with the other end first, with $u_0 \neq u_0'$. In both cases, the potential is periodic with period b, the distance between nucleotides. This corresponds to an attractive potential when the polymer enters the pore with the one end first and a repulsive potential when it enters with the other end first.

In Figure 6 we show our results of the transit time distribution, for the model with free energy $W = W_0^{'} + W_1 + u(x)$, where u(x) is the interaction between the polymer and the pore, explained above, obtained using 100000 polymers each with 210 nucleotides. We have used here $l^{'} = 3$ in $W_0^{'}$, and $u_0 = 4.5 k_B T$ , $u_0' = 0$ in u(x). With these parameters we clearly obtain three peaks in the distribution of lifetimes. The results also look much more like the experimental data of KBBD. By comparing with Fig. 4, we recognize that the third peak, at lifetime of $11 t_0$ corresponds to the second peak, also at lifetime of $11t_0$, in the case of no interaction between polymer and pore. The second peak which is the highest peak here, seems to be created by interaction of the polymer with the pore. If we identify position of the third peak at $11t_0$ with that of the third peak at $1400ms$ of the experimental data of KBBD, we obtain $t_0 \approx 127ms$.

## III. CONCLUSION AND DISCUSSION

We have simulated the translocation of a polymer through a nanopore, driven by an external electric field, using the continuous time Monte Carlo method. The nanopore is small enough so that only single strands of the polymer can pass through. We consider separately two models of interaction with the external electric field. In the first case the electric field gives a constant pull on the polymer. When there is no interaction of the polymer with the pore, the transit time distribution consists of a peak at small transit times corresponding to polymers partially entering the pore but then retracted back into the cis side. This peak is independent of the size of the polymer. The second peak at larger transit time corresponds to the polymer passing completely through the pore. Its shape is that of a Gaussian and the position of this peak increases proportional to the size of the polymer. The width of this peak is proportional to the square root of the average transit time $t_p$, or the square root of the polymer size. These results are in agreement with the results of the Langevin equation corresponding to the model studied by Lubensky and Nelson. However our results are different from their results obtained using a different method to calculate the first passage time distribution. An interaction between the pore and the polymer can be added in the form of an asymmetric saw-tooth potential suggested by Lubensky and Nelson, characterized by the parameters $u_0$ which is the height of the potential and an asymmetry parameter $a$ . The asymmetry parameter corresponds to the polymer interacting differently with the pore when it enters the pore with one end first than with the other end first. With this interaction one obtains three peaks in the transit time distribution just as in the experiment of KBBD, but the shape of the distribution is very different.

We also studied another model in which the pull of the external electric field on the polymers at the pore increases with the length of polymer transmitted through the pore. This is because as the polymer get pulled through the pore, more charges will be on the trans side. This gives a stronger force in the electric field. The force due to the electric field on the cis side is assumed to be cancelled by the reaction of the membrane. When there is no interaction between the polymer and the pore, a Langevin equation can be derived for the time development of the length of polymer having passed through the pore at time t, depending on a characteristic time $t_f$ of the electric field. The solution of this equation shows that the transit time $t_p$ is not Gaussian distributed due to its nonlinear logarithmic dependence on the random noise. Its dependence on the size of the polymer is at most $\log N$ and its dependence on the characteristic time $t_f$ is linear for small N. Our simulations show that the transit time $t_p$ is indeed not Gaussian distributed and its dependence on the characteristic time $t_f$ is linear for small N, but its dependence on the size of the polymer is weaker than $\log N$ and actually saturates at large N. Including the asymmetric interaction of the polymer with the pore results in three peaks in the distribution of transit times and the distribution itself is much more like that in the experiment of KBBD, except that our first peak is lower. The first peak in the distribution, which corresponds to polymers partially entering and then retracting from the pore from the cis side, is much lower compared to experimental data. In our calculation, we have assumed that the polymer is always a single strand in the cis side to start with. Experimentally, some of the polymers could form partially double strands. These double stranded polymers could not pass through the pore due to their size and could actually jam the pore. In the experiment, in order to clear the jamming, the voltage had to be reversed. Beside jamming, which must be cleared by reversing the voltage, these double stranded polymers, since they are physically too large to pass through, must also lead to increase number of retractions, which can explain the increased first peak seen in the experiment.

We have compared our simulation only with the model of Lubensky and Nelson. The reason is that although many published simulations of the polynucleotide translocation exist, many of which were quoted in our references, to the best of our knowledge, there is no simulation of the model proposed by Lubensky and Nelson. We believe our work is the first simulation of this model. In fact, as far as we know, there are no simulations that produce the three peaks in the experiment of Kasianowicz, Brandin, Branton and Deamer (KBBD). Even in the theory of Lubensky and Nelson, it is only predicted that three peaks should be seen using an asymmetric interaction with the pore. But only the case of one peak with no interaction with the pore was actually calculated. Our simulation is the first time in which three peaks are actually produced. For the same reason we have only compared our simulation with the experimental data of KBBD.

We have assumed, as well as Lubensky and Nelson, an asymmetric interaction of the polymer with the pore. This is a reasonable assumption. The value 4.5kT that we have chosen is of no particular significance. It is only a parameter value for the interaction that seems to give the best agreement with experimental data. We happened to show our result at this value of the parameter. A smaller value would have also given reasonable, even though not as good agreement. Also the location of the first peak in the translocation time (at very short times) could have been biased in the experiments due to limited system response as explained in KBBD paper. Just what is the interaction of the polymer with the pore? This question can only be answered by future experiments.

In this paper we have only compared with the experimental results of KBBD. There are many new experimental results on DNA translocation since KBBD. However, they deal mainly with others aspects of the DNA translocation problem than the distribution of passage times. Storm et al. [14] studied the power scaling of translocation times versus length, using solid-state nanopore. Akimentiev et al. [11] studied DNA translocation as a new technique for sequencing DNA. Chang et al. [19] studied the fluctuations in ionic current during DNA translocation through nanopore. They are not directly relevant to the problem in the present paper. For instance, we are not aware of any other experimental result that produces the three peaks in the distribution of passage times, besides that of KBBD.

Recently Mathe et al [20] experimentally studied the orientation discrimination of single-stranded DNA inside the alpha-hemolysin nanopore. They found that the DNA-channel interactions depend strongly on the orientation of the ssDNA molecule with respect to the pore, both in voltage driven and in zero voltage diffusions through the pore. Taking advantage of the finding that ssDNA can enter the pore but double-stranded DNA cannot, they used DNA haripin molecules with a long single-stranded overhang which can be either a 3' end or a 5' end. In this way they could determine precisely with which end the ssDNA molecules entered the pore. The resulting current histogram which is proportional to the distribution of translocation times exhibits two well defined peaks which can be well fit by a double Gaussian distribution. The Gaussian distribution of the translocation times is in agreement with our result, Eqn. (14). This is different from the KBBD result which shows three peaks. But, as mentioned above, the position of the short life peak in the KBBD experiment is sensitive to the bandwidth in the experiment, in contrast to the other two long life time peaks whose positions are independent of the bandwidth. In this sense our fit to the short time peak is therefore probably coincidental. In addition Mathe et al clarified the origin of this asymmetry using molecular dynamic simulation. In a confined pore, the ssDNA straightens and its bases tilt towards the 5' end, assuming an asymmetric conformation. As a result, the bases of a 5'-threaded DNA experience larger effective friction.

*Email: puiman_lam@subr.edu

Acknowledgement. The research of PML is supported by the Petroleum Research Fund of the American Chemical Society. He would like to thank the Institute of Theoretical Physics, Chinese Academy of Sciences, for hospitality, where part of this work was completed.

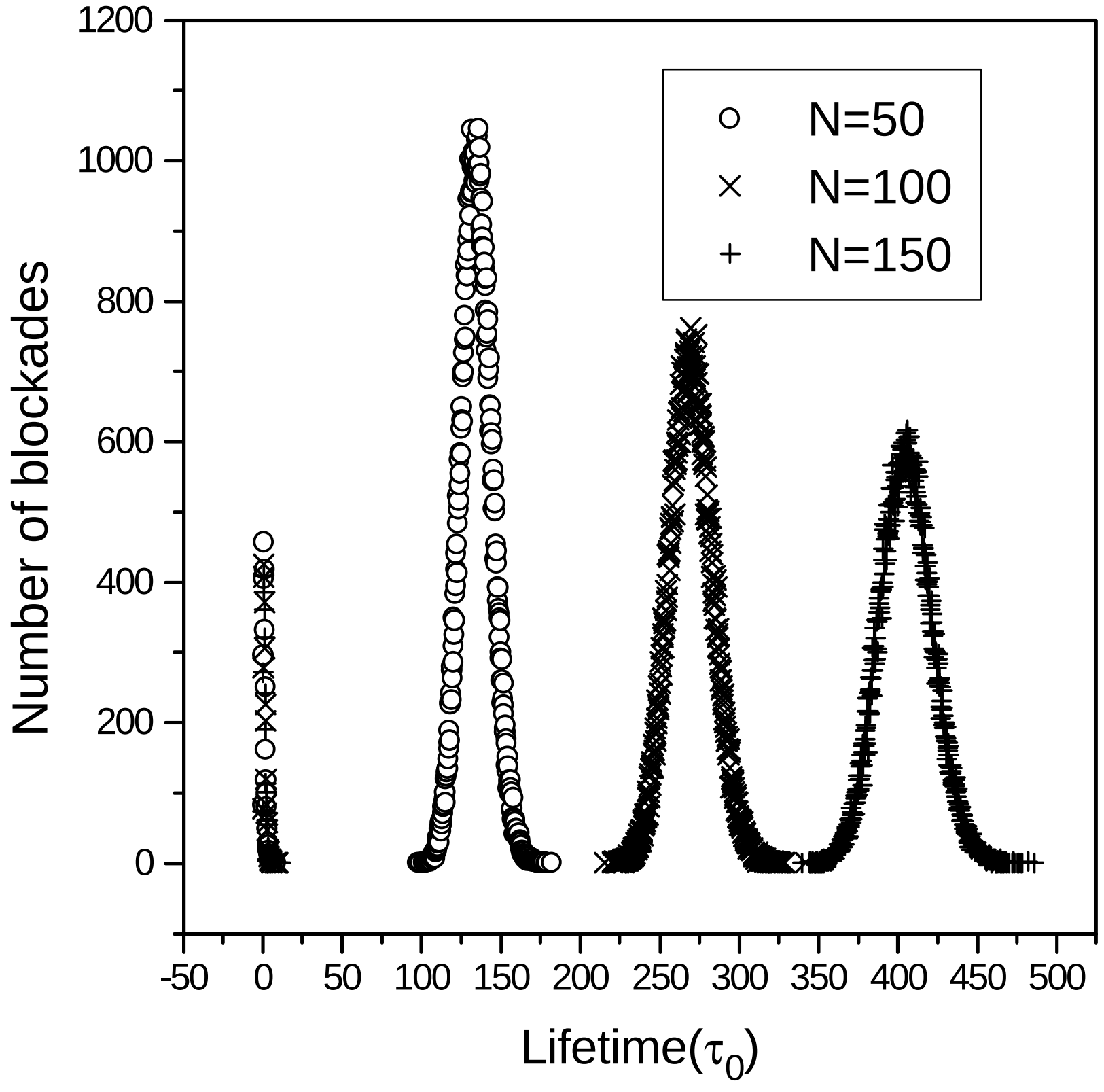

Figure 1a

Figure 1a: Distribution of passage times for different lengths Nb of the polymer, for the case of no interaction between polymer and the pore. Free energy $W=W_0$.

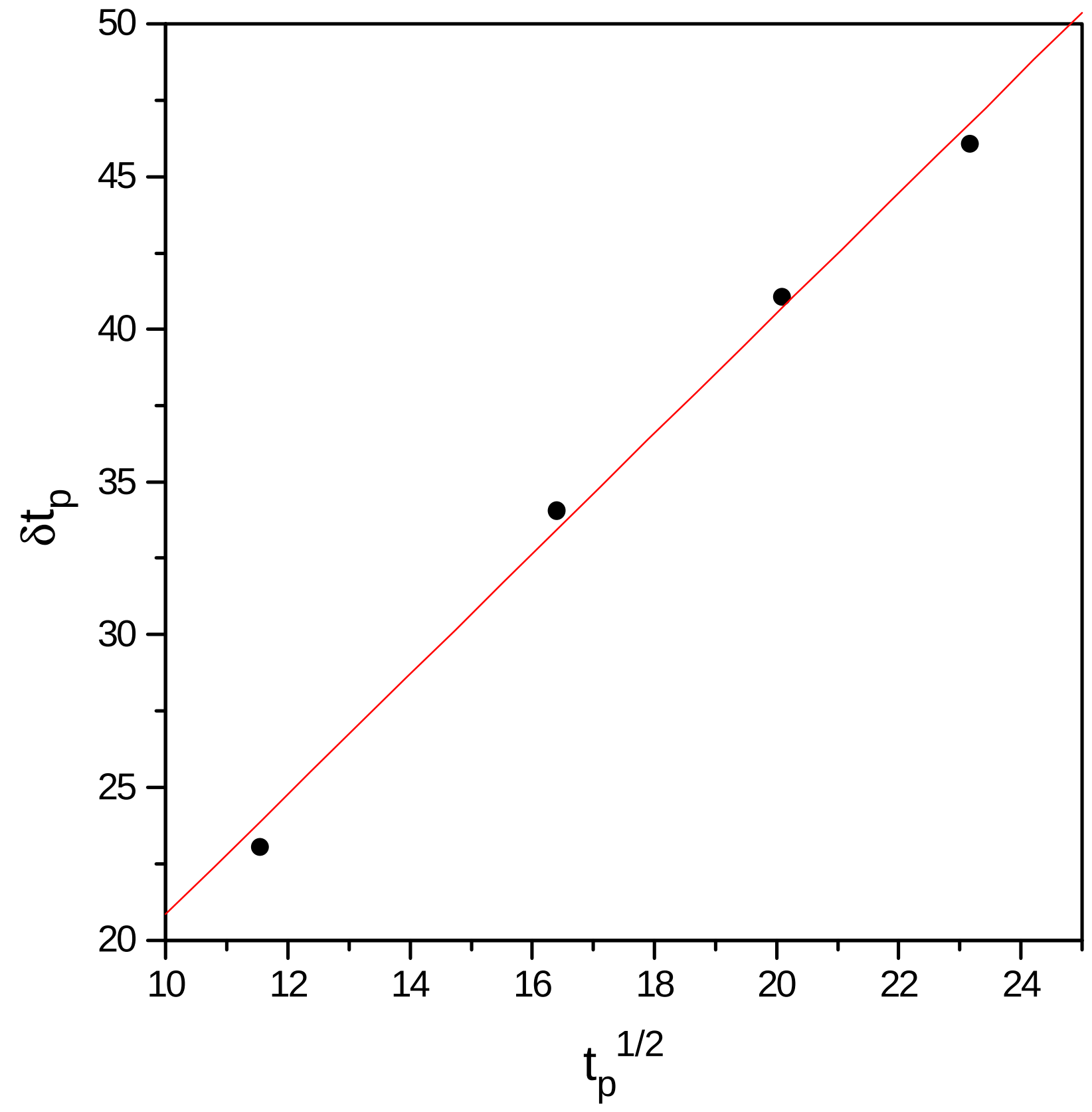

Figure 1b

Figure 1b: Average fluctuations $dt_p$ of the transit time $t_p$ versus $t_p^{1/2}$.

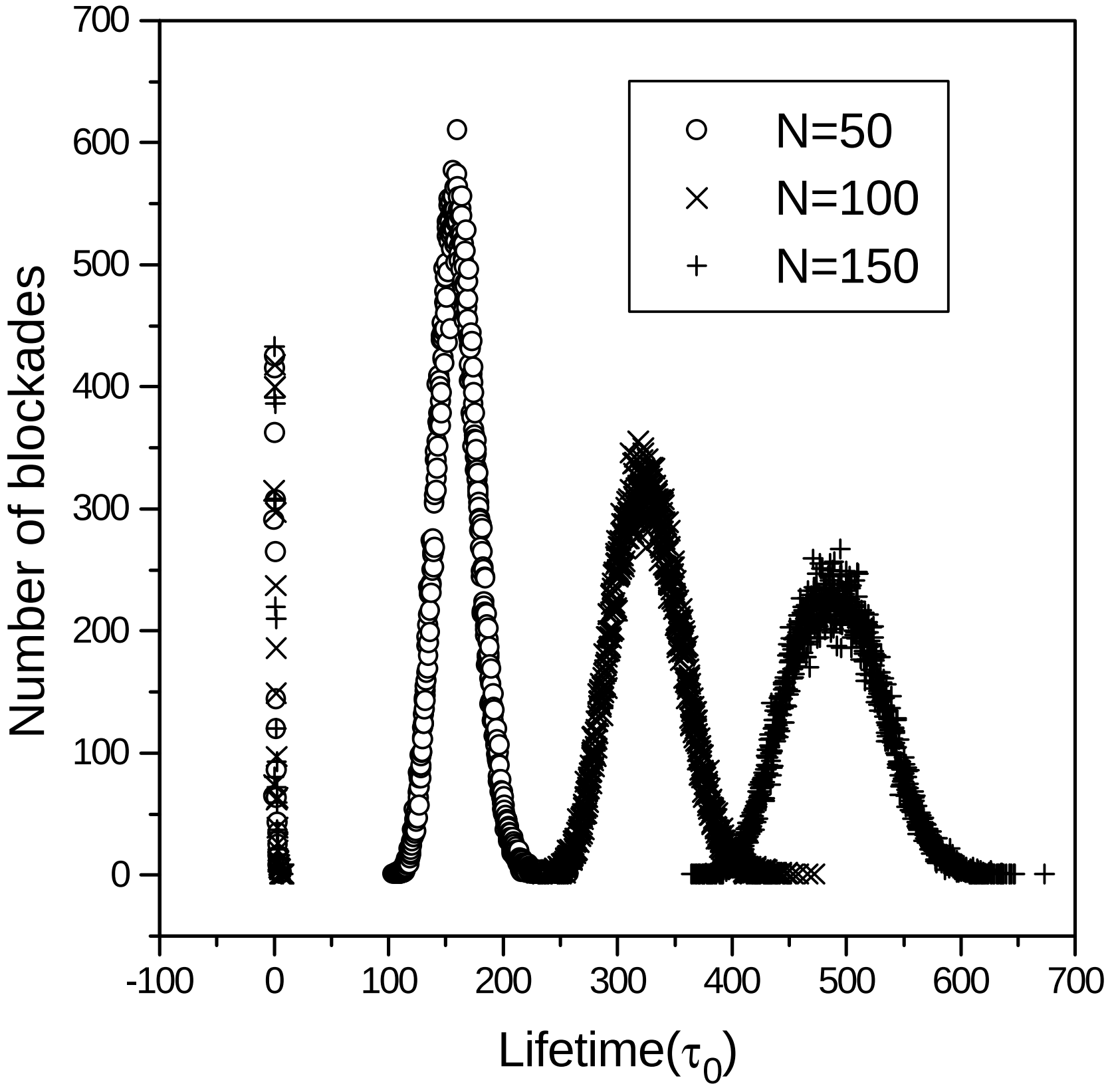

Figure 2

Figure 2 : Distribution of passage times for different lengths Nb of the polymer, for the case of no interaction between polymer and the pore. Free energy $W=W_0+W_1$.

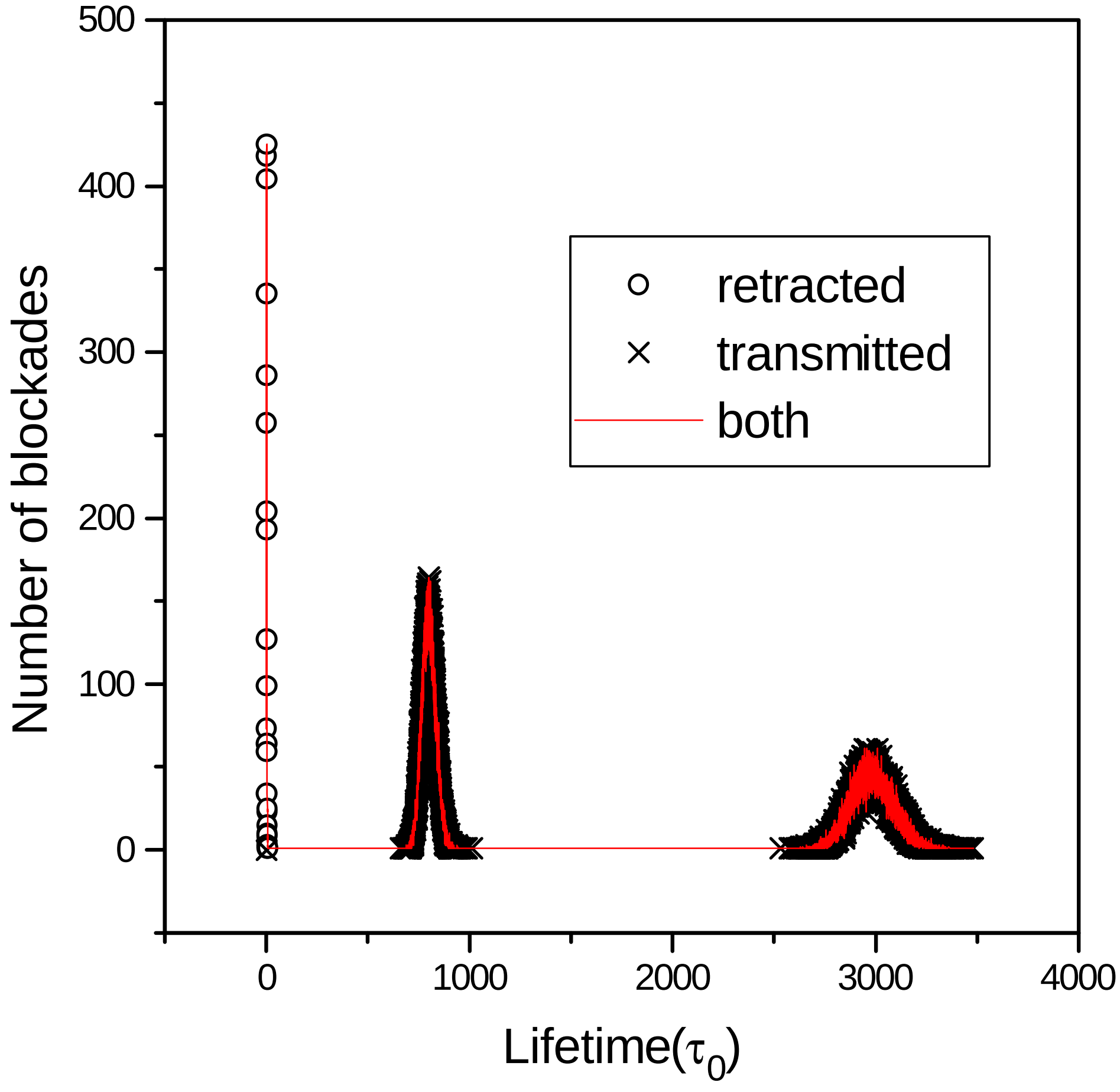

Figure 3.

Figure 3: Distribution of passage times for the case with interaction between polymer and the pore: $W = W_0 + W_1 + \frac{u(x)}{k_B T}$, with $l = 5$ in $W_0$, $u_0 = 0.2 k_B T$ and $a = 0.1$ in u(x).

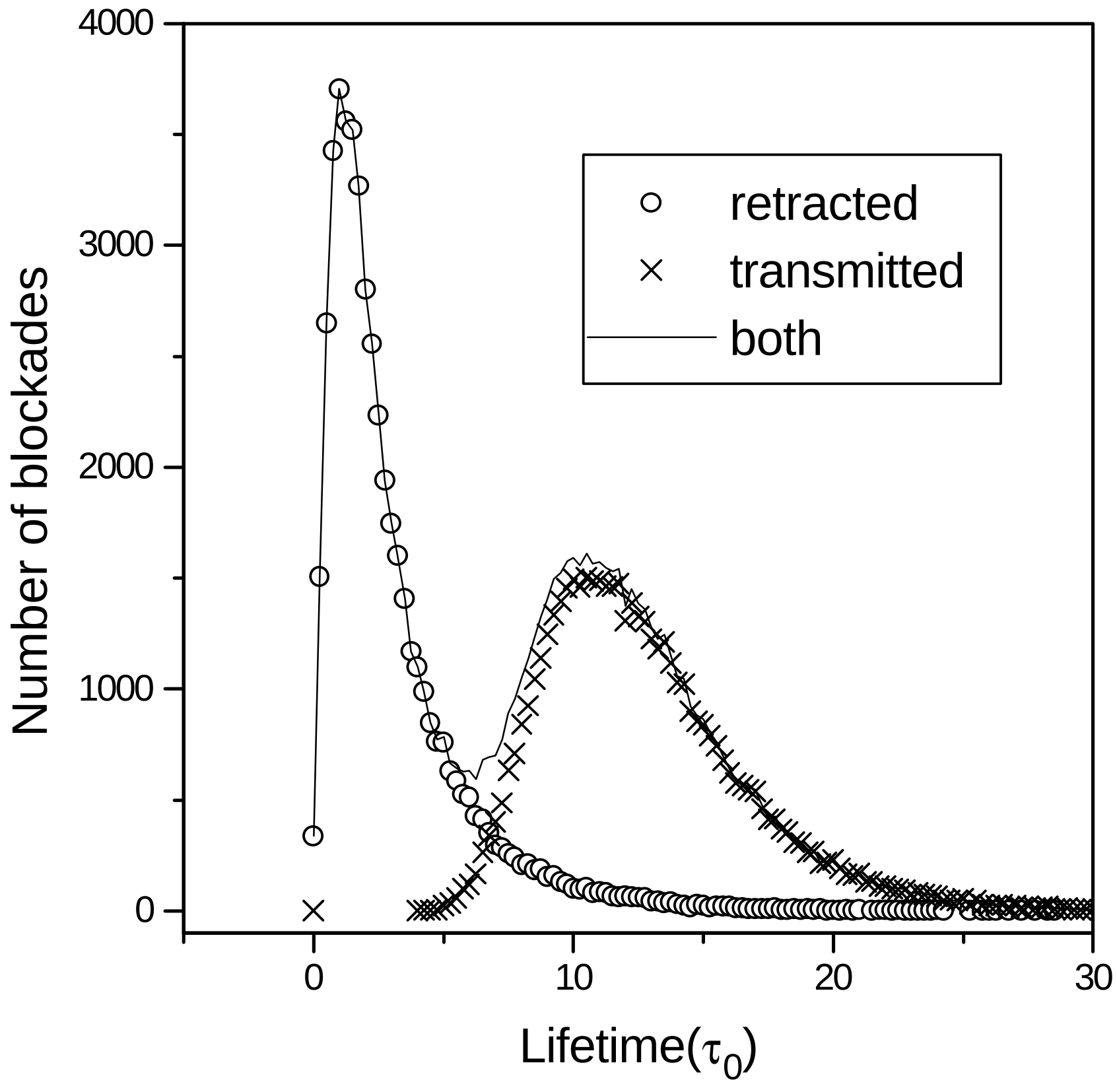

Figure 4

Figure 4: The distribution of transit times for free energy $W = W_0{}'$, with $l' = 3$, and N=200

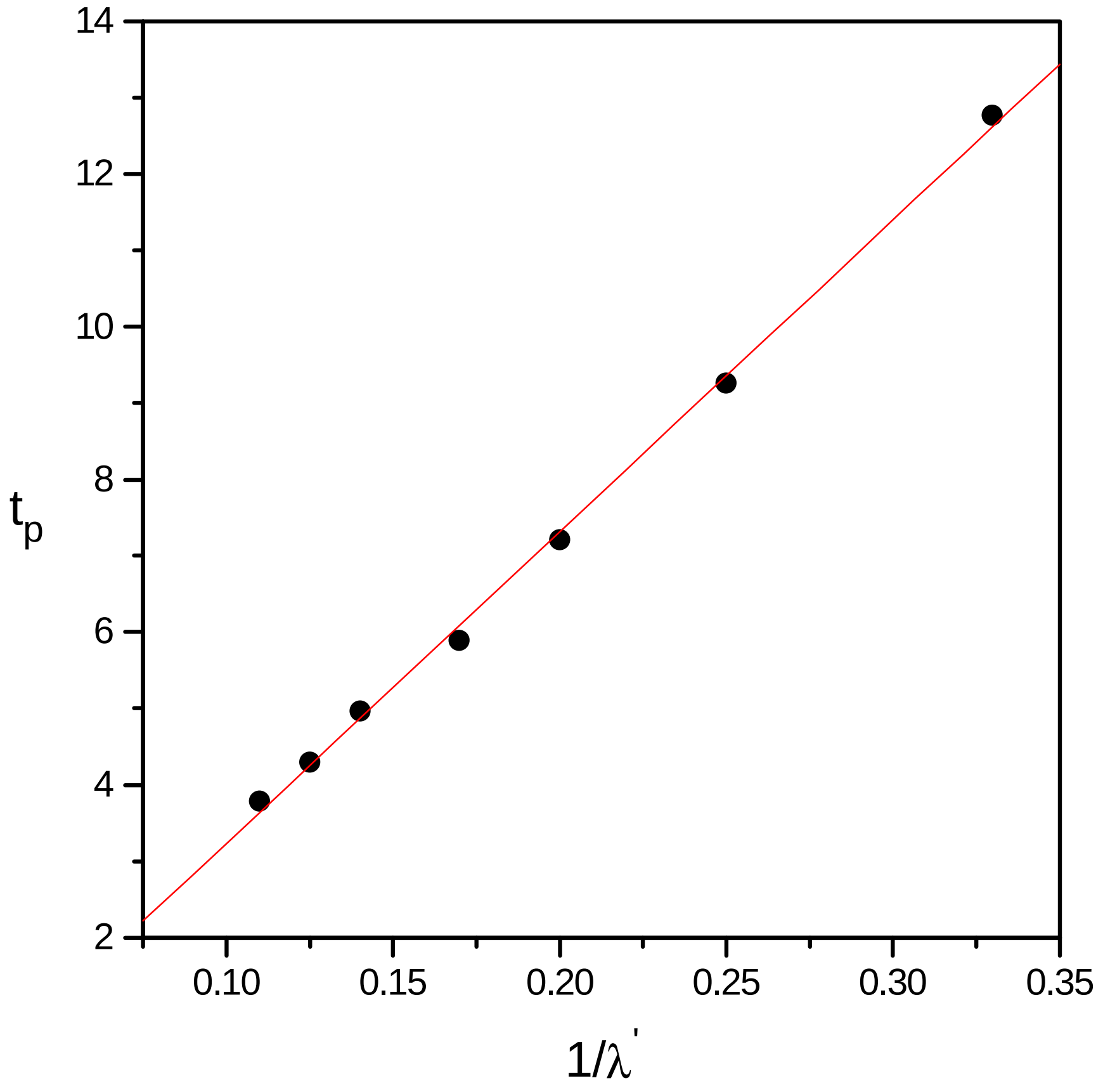

Figure 5

Figure 5: Average transit time $t_p$ in the model with free energy $W = W_0{}'$, versus $1/\, l'$ which is proportional to characteristic time $t_f$ due to the external field, for polymer chain length N=20.

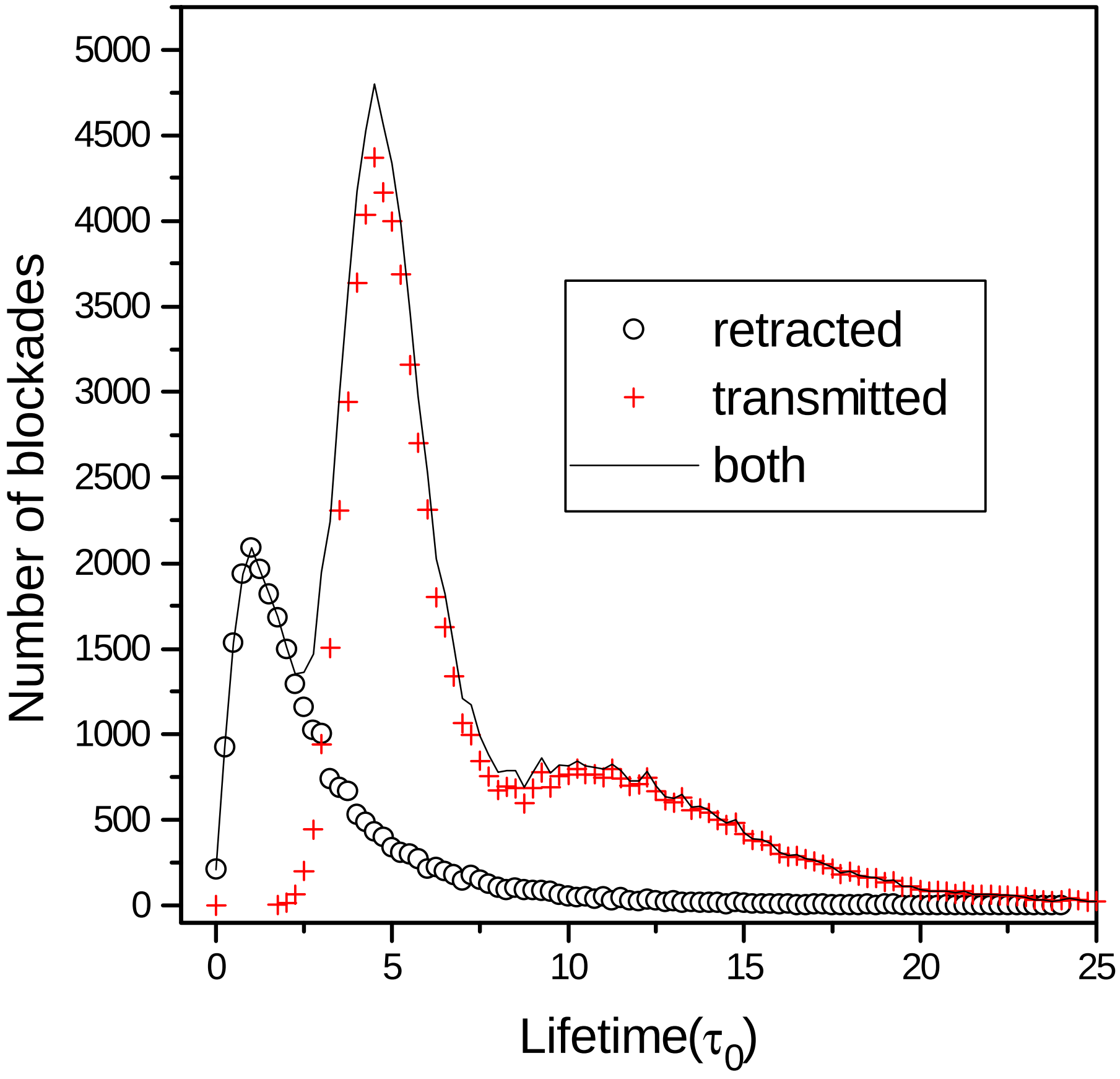

Figure 6

Figure 6: Transit time distribution in the model $W(x) = W_0{}'(x) + W_1(x) + u(x)$ where u(x) is the interaction between the polymer and the pore, as explained in the text. We have chosen $l' = 3$ in $W_0{}'(x)$, and $u_0 = 4.5 k_B T$, $u_0{}' = 0$ in u(x).